\documentstyle[11pt,nstar,psfig]{article}

\begin{document}

\vspace*{-2.0cm}
{\bf\hfill\large MKPH-T-98-4}
\vspace*{0.5cm}

\title{Eta Photoproduction}
\author{Lothar Tiator\thanks{E-mail: 
{\tt tiator@kph.uni-mainz.de}}
and Germar Kn\"ochlein\thanks{E-mail: 
{\tt knoechle@kph.uni-mainz.de}}\\
{\em Institut f\"ur Kernphysik,
Johannes Gutenberg-Universit\"at,
D-55099 Mainz, Germany}\\
\vspace{0.3cm}
Cornelius Bennhold\thanks{E-mail: 
{\tt bennhold@gwis2.circ.gwu.edu}}\\
{\em Center for Nuclear Studies, Department of Physics, The George
Washington University, Washington, D.C., 20052}}

\maketitle

\begin{abstract}
  We present a combined analysis of the new eta photoproduction data
  for total and differential cross sections, target asymmetry and
  photon asymmetry. Using basic assumptions, this allows a
  model-independent extraction of the $E_{2-}$ and $M_{2-}$ multipoles
  as well as resonance parameters of the $D_{13}(1520)$ state.  At
  higher energy, we show that the photon asymmetry is extremely
  sensitive to small multipoles that are excited by photons in the
  helicity $3/2$ state. These could be, e.g., the $F_{15}(1680)$, the
  $F_{17}(1990)$, or the $G_{17}(2190)$ resonances.
\end{abstract}

\section*{INTRODUCTION}
Over the last several years, eta photoproduction has demonstrated its
potential as a new, powerful tool to selectively probe certain
resonances that are difficult to explore with pions. It is well known
that the low energy behavior of the eta production process is governed
by the $S_{11}(1535)$ resonance\cite{Benn91,Tiat94,Benm95}. The
recent, precise measurements of total and differential cross sections
for eta photoproduction at low energies\cite{Krus95,Wilh93} have
allowed determining the $S_{11}(1535)$ resonance parameters with
unprecedented precision. However, it is because of the overwhelming
dominance of the $S_{11}$ that the influence of other resonances in
the same energy regime, such as the $D_{13}(1520)$, is difficult to
discern.  It has been pointed out\cite{Tiat94} that polarization
observables provide a new doorway to access these non-dominant
resonances which relies on using the dominant $E_{0+}$ multipole to
interfere with a smaller multipole. Especially the polarized photon
asymmetry was shown to be sensitive to the $D_{13}(1520)$. Recently,
polarization data for the target and photon asymmetries in eta
photoproduction were measured at ELSA\cite{Bock98} and
GRAAL\cite{Hour98}, respectively, for the first time.  Taken together
with the data for the unpolarized cross section from MAMI, they allow
a determination of the $D_{13}(1520)$ contribution in eta
photoproduction.

\section*{MULTIPOLE ANALYSIS}

In the following all considerations refer to the c.m. frame.  
The three measured observables
have the following representation in terms of the response functions
defined in \cite{Knoe95}:
\begin{eqnarray}
\frac{d \sigma}{d \Omega}& = &\frac{|\vec k_{\eta}|}{|\vec q|}
R_T^{00}\, ,\\
T & = & \frac{R_T^{0y}}{R_T^{00}}\, ,\\
\Sigma & = & - \frac{^{c}R_{TT}^{00}}{R_T^{00}}\, .
\end{eqnarray}

Because of the overwhelming dominance of the $S_{11}$ channel in eta
photoproduction, the observables can be expressed in terms of $s$--wave 
multipoles and interferences of the $s$ wave with other
multipoles. In the CGLN basis this leads to an $F_1$ dominance and
the observables can simply be expressed as
\begin{eqnarray}
\label{cgln}
R_T^{00} & = & |F_1|^2 - \mbox{Re} \left\{ 2 \cos\theta F_1^* F_2 -
    \sin^2\theta F_1^*F_4 \right\},\\
R_T^{0y} & = & 3 \sin\theta\, \mbox{Im} \left\{ F_1^* F_3 +
    \cos\theta F_1^*F_4 \right\},\\
^cR_{TT}^{00} & = & \mbox{Re} \left\{F_1^* F_4 \right\}.
\end{eqnarray}
If we retain only interferences with $p$-- and $d$--waves (an
approximation that is valid at least up to 1 GeV photon lab energy)
we obtain
\begin{eqnarray}
\label{o1}
R_T^{00} & = &
|E_{0+}|^2 - \mbox{Re} \left[ E_{0+}^* \left( E_{2-} - 3 M_{2-} \right) \right] 
\nonumber \\
& & + 2 \cos \theta \mbox{Re} \left[ E_{0+}^* 
\left( 3 E_{1+} + M_{1+} - M_{1-} \right) \right]
\nonumber \\
& & + 3 \cos^2 \theta \mbox{Re} \left[ E_{0+}^* \left( E_{2-} - 3 M_{2-} \right) 
\right]\, ,\\
\label{o2}
R_T^{0y} & = &
3 \sin \theta \mbox{Im} \left[ E_{0+}^* \left( E_{1+} - M_{1+} \right) \right]
\nonumber \\
& & - 3 \sin \theta \cos \theta 
\mbox{Im} \left[ E_{0+}^* \left( E_{2-} + M_{2-} \right) \right]\, ,\\
\label{o3}
^{c}R_{TT}^{00}& = & 
- 3 \sin^2 \theta \mbox{Re} \left[ E_{0+}^* \left( E_{2-} + M_{2-} \right) \right]
\, .
\end{eqnarray}
With the following angle-independent quantities
\begin{eqnarray}
a & = & |E_{0+}|^2- \mbox{Re} \left[ E_{0+}^*\left( E_{2-}-3 M_{2-} \right) \right]
\, ,\\
b & = & 2 \mbox{Re} \left[ E_{0+}^* \left( 3 E_{1+} + M_{1+} - M_{1-} \right) \right]
\, ,\\
c & = & 3 \mbox{Re} \left[ E_{0+}^* \left( E_{2-} - 3 M_{2-} \right) \right]\, ,\\
d & = & 
\frac{1}{a + \frac{1}{3} c}
3 \mbox{Im} \left[ E_{0+}^* \left( E_{1+} - M_{1+} \right) \right]\, ,\\
e & = & - 3
\frac{1}{a + \frac{1}{3} c}
\mbox{Im} \left[ E_{0+}^* \left( E_{2-} + M_{2-} \right) \right]\, ,\\
f & = & 3 \frac{1}{a + \frac{1}{3} c}
\mbox{Re} \left[ E_{0+}^* \left( E_{2-} + M_{2-} \right) \right]\, ,
\end{eqnarray}
we can express the observables in a series of $\cos\theta$ terms that
can be fitted to the experimental
data at various energies $E_{\gamma,lab}$
% as a function of $E_{\gamma,lab}$
\begin{eqnarray}
\frac{d \sigma}{d \Omega}& = & \frac{|\vec k_{\eta}|}{|\vec q|}
\left(a + b \cos \theta + c \cos^2 \theta\right)\, ,\\
T & = & \sin \theta \left(d + e \cos \theta \right)\, ,\\
\Sigma & = & f \sin^2 \theta \, .
\end{eqnarray}
It is remarkable that a combined analysis of the three above
observables allows a determination of the $d$--wave contributions
to eta photoproduction once the quantities $a$, $c$, $e$ and $f$ have
been determined from experiment.  
Already with the knowledge of $e$ and $f$ the helicity $3/2$ multipole
$B_{2-}$, defined below, and the phase relative to the $S_{11}$ channel can be
determined:
\begin{eqnarray}
| B_{2-} | \equiv | E_{2-} + M_{2-} | & = & \frac{\sqrt{e^2+f^2}}
{3\sqrt{a+c/3}}\, ,\\
\tan (\phi_{E_{0+}}-\phi_{B_{2-}}) & = & \frac{e}{f}\, .
\end{eqnarray}
If one neglects electromagnetic effects from the background of
eta photoproduction 
affecting the phase of the electric and magnetic multipoles differently
($\phi_{E_{l \pm}} = \phi_{M_{l \pm}} = \phi_{l \pm}$), one can write
\begin{eqnarray}
E_{l \pm} & = | E_{l \pm} | e^{i \phi_{l \pm}} \, ,\\
M_{l \pm} & = | M_{l \pm} | e^{i \phi_{l \pm}}\, , 
\end{eqnarray}
and one finds the following representation for the real and imaginary parts 
of the $d$--wave multipoles:
\begin{eqnarray}
\mbox{Re} E_{2-} & = & \frac{1}{4} \sqrt{a + \frac{1}{3} c} 
\left(f \cos \phi_{0+} + e \sin \phi_{0+} \right)
\left(1 + \frac{c}{3 f}\right)\, ,\\
\mbox{Im} E_{2-} & = & \frac{1}{4} \sqrt{a + \frac{1}{3} c} 
\left(f \sin \phi_{0+} - e \cos \phi_{0+} \right)
\left(1 + \frac{c}{3 f}\right)\, ,\\
\mbox{Re} M_{2-} & = & \frac{1}{12} \sqrt{a + \frac{1}{3} c} 
\left(f \cos \phi_{0+} + e \sin \phi_{0+} \right)
\left(1 - \frac{c}{f}\right)\, ,\\
\mbox{Im} M_{2-} & = & \frac{1}{12} \sqrt{a + \frac{1}{3} c} 
\left(f \sin \phi_{0+} - e \cos \phi_{0+} \right)
\left(1 - \frac{c}{f}\right)\, .
\end{eqnarray}
We note that this determination of the $E_{2-}$ and $M_{2-}$
multipoles is rather model independent. To be more explicit we list
the assumptions used to arrive at the above formulae:
\begin{itemize}
\item Phase difference between electric and
  magnetic multipoles neglected, $\phi_{E_{l \pm}} = \phi_{M_{l \pm}}
= \phi_{l \pm}$ 
\item Restriction to the truncated multipole representation of Eqs.\
  (\ref{o1}), 
  (\ref{o2}), 
  (\ref{o3}) 
\item Knowledge of the phase of the $E_{0+}$ multipole.
\end{itemize}
The last point deserves further discussion: From total cross
section data \cite{Krus95} it is obvious that in the region of the
$S_{11}(1535)$ resonance the cross section can be perfectly fitted by
a Breit--Wigner resonance resulting in $s$--wave dominated
differential cross sections. An investigation of the background from
the Born terms \cite{Tiat94} yielded a very small eta--nucleon
coupling constant. As a consequence, the $E_{0+}$ multipole can be treated as
being completely dominated by the $S_{11}(1535)$ contribution, which,
as shown in ref.\cite{Krus95}, allows
parametrizing it through a Breit--Wigner form. 
In principle, an
arbitrary phase for the complex $E_{0+}$ multipole could be added
which is set equal to 0 by convention. 
For the complex $E_{0+}$ multipole we use the Breit--Wigner
parametrization 
\begin{equation}
E_{0+} = - \sqrt{\frac{a}{4 \pi}} \frac{\Gamma^* M^*}{{M^*}^2 - W^2 -
  i M^* \Gamma(W)}\, ,
\end{equation}
where $W$ is the c.m. energy. 
The energy dependence of the resonance width is given by 
\begin{equation}
\Gamma(W) = \Gamma^* \left( 
b_{\eta} \frac{|\vec k|}{|\vec k^*|}
+ b_{\pi} \frac{| \vec k_{\pi}|}{|\vec k_{\pi}^*|} + b_{\pi\pi}
\right) \, .
\end{equation}
The analysis of the $E_{0+}$
interference with the $E_{2-}$ and $M_{2-}$ multipoles determines
the $d$ wave multipoles and therefore the difference $\phi_{2-}
- \phi_{0+}$. It does not yield direct information
on $\phi_{2-}$. However, making the above assumptions for the
$E_{0+}$ multipole and thus the phase $\phi_{0+}$ permits the
determination of $\phi_{2-}$. 

To perform a similar analysis of the $p$--wave multipoles more
information from additional polarization observables is required; in
particular, a measurement of the recoil polarization would be very
helpful. As before we obtain 
\begin{eqnarray} 
  P & = & \frac{R_T^{y0}}{R_T^{00}}\, ,\\ & = & \sin \theta \left(g +
    h \cos \theta \right)\, \\ 
\end{eqnarray}
with
\begin{eqnarray}
g & = & - \frac{1}{a + \frac{1}{3} c} \mbox{Im} 
\left[ E_{0+}^* \left( 2 M_{1-} + 3 E_{1+} + M_{1+} \right) \right]\, ,\\
h & = & 3 \frac{1}{a + \frac{1}{3} c}
\mbox{Im} \left[ E_{0+}^* \left( E_{2-} - 3 M_{2-} \right) \right]\, .
\end{eqnarray}
After performing single-energy fits we used
a polynomial fit to the energy dependence of the coefficients 
$a$, $b$, $c$, $d$, $e$ and $f$ in order to arrive at a global (energy
dependent) solution for the multipoles. This has several advantages:
First the experimental data have been obtained in different set--ups at
different labs, thus their energy bins do not match each other. Second,
except for quantity $a$ that is in principle determined already by the
total cross section, all other quantities contain considerable error
bars, therefore, a combined fit can reduce the uncertainty of individual
measurements considerably. In a simple Taylor expansion in terms of
the eta momentum with only 1-3 parameters in each coefficient we obtain
good results for an energy region from threshold up to about 900 MeV.

\section*{RESULTS}

Fig. 1 shows 4 out of 10 angular distributions measured by the
TAPS collaboration at Mainz \cite{Krus95} in the energy range between
716 and 790 MeV. While our isobar model falls a bit low close
to threshold, a perfect fit is possible using the Ansatz in
Eq. (16). Our results for the coefficients $a$, $b$ and $c$ agree
perfectly with the results ontained in Ref.~\cite{Krus95}.
As mentioned before, the $a$ coefficient can be fitted
to a Breit-Wigner form with an energy-dependent width leading, e.g., to
parameters of $M^*=(1549\pm 8)MeV$, $\Gamma_R=(202\pm 35) MeV$ and an
absolute value of the $s$-wave multipole at threshold, $|E_{0+}|=16.14
\cdot 10^{-3}/m_\pi^+$ (Fit 1, Ref.~\cite{Krus95}). For our purpose
here it is more convenient to use a general polynomial expansion 
as mentioned above.
\begin{figure}[htbp]
\centerline{\psfig{file=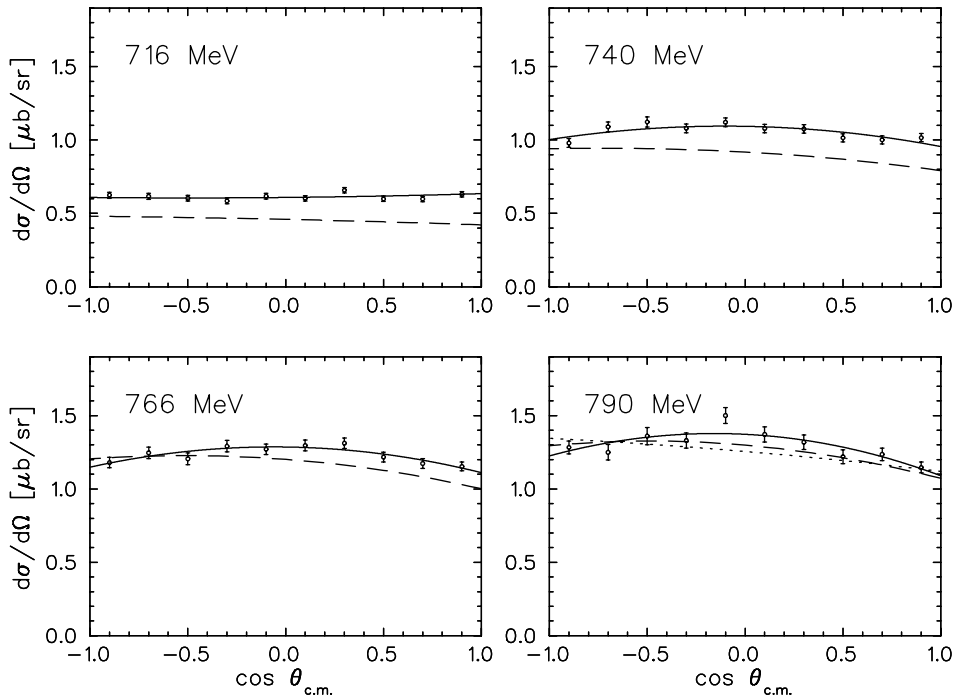,width=12.5cm,angle=90,silent=}}
\caption[thanks]{\label{fig1} Differential cross section for 
$p(\gamma,\eta)p$. The solid lines show the fit to the experimental
data of Krusche et al. \cite{Krus95}. The dashed lines show our 
calculations in the isobar model \cite{Knoe95}. The dotted line at the
highest photon lab energy of $790 MeV$ are obtained from our
calculations when the $D_{13}$ resonance is turned off.
}
\end{figure}

Fig. 2 shows the target polarization with the preliminary data from
Bonn\cite{Bock98}. Here our isobar model fails to reproduce the
angular shape of the data. In particular there is no node in our
calculation and the role of the $D_{13}$ resonance plays a very small
and insignificant role. In our previous coupled channel analysis the
$D_{13}$ resonance came out much stronger and a node developed,
however, with a minus sign at forward and a positive sign at backward
angles. This is opposite to the experimental observation and, as we
will see later, indicates a drastically different relative phase
between $s$- and $d$--waves. With the ansatz of Eq.~(17) we can
fit the data and obtain a node at low energies that disappears around
800 MeV. 
\begin{figure}[htbp]
\centerline{\psfig{file=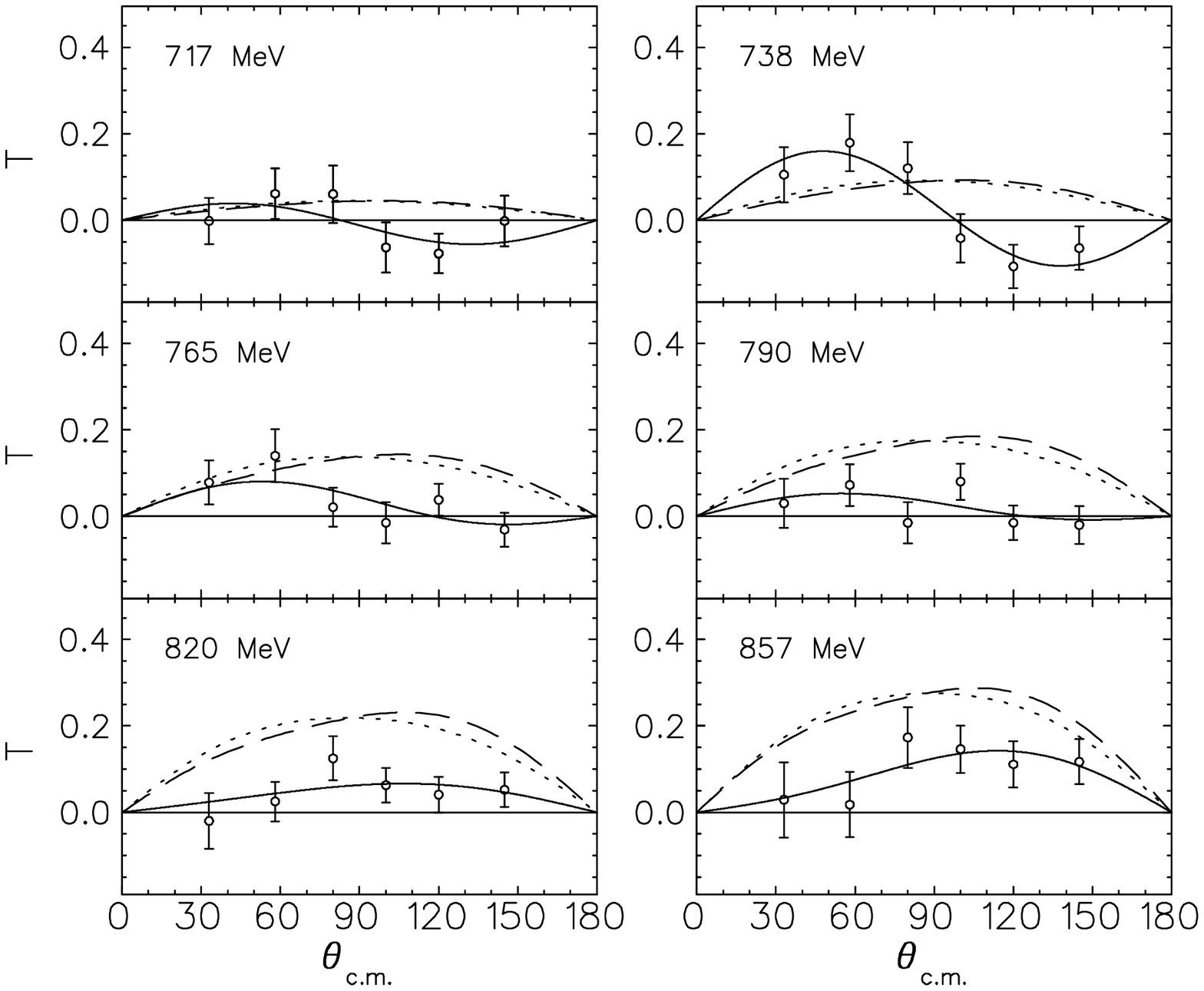,width=12cm,silent=}}
\caption[thanks]{\label{fig2} Target asymmetry for
  $p(\gamma,\eta)p$. The dashed and dotted lines
  show our calculations in the isobar model \protect\cite{Knoe95} with
  and without the $D_{13}(1520)$ resonance. The solid line is the
  result of our fit to the experimental data of \protect\cite{Bock98}.
  }
\end{figure}

In Fig. 3 we show our isobar calculations for the photon asymmetry.
This observable has been measured recently at GRAAL \cite{Hour98},
however, the data are still in the analysis. A preliminary comparison,
however, shows general agreement for energies below 1~GeV. From our
calculations the importance of the $D_{13}$ channel for the photon
asymmetry becomes obvious. Without this nucleon resonance, the
asymmetry would be almost zero up to about 900~MeV. Even as the
experimental data for the photon asymmetry are not yet available we can
already perform a preliminary analysis of the $D_{13}$ multipoles
under the constraint of the photon asymmetries determined by our
isobar model. In this case, all coefficients of Eqs.~(10-15) are
available and we can evaluate the $d$--wave multipoles using
Eqs.~(21-24). As mentioned before, the solution for the individual
multipoles $E_{2-}$ and $M_{2-}$ requires the additional assumption for
the phase of the $s$--wave amplitude. This is taken from the
Breit-Wigner Ansatz Eqs.~(27-28) with the parameters of fit 1 in
Ref.~\cite{Krus95}. Of course, this form is rather ad hoc, however,
comparing with coupled channels calculations \cite{Kais97,Feus97}
we find that the results of these very different 
approaches agree very well not only for the absolute
magnitude of the $s$--wave but also for the phase.
\begin{figure}[htbp]
\centerline{\psfig{file=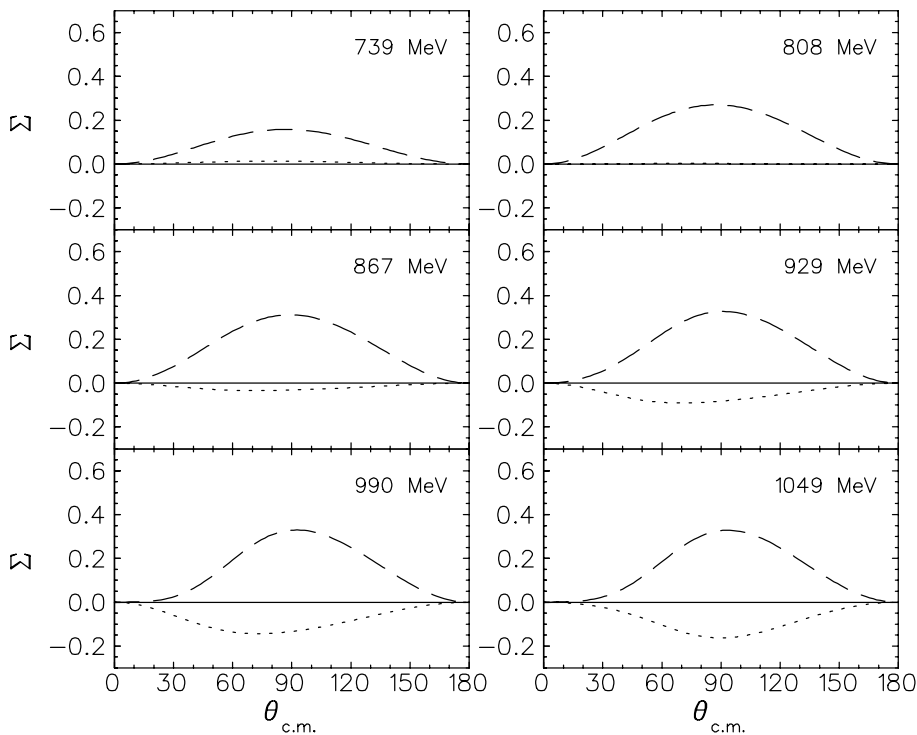,width=12cm,silent=}}
\caption[thanks]{\label{fig3} Photon
  asymmetry for $p(\gamma,\eta)p$. The dashed and dotted lines
  show our calculations in the isobar model \protect\cite{Knoe95} with
  and without the $D_{13}(1520)$ resonance.
}
\end{figure}

Fig. 4 shows the result of our multipole analysis and compares it
with our isobar model calculation. The biggest difference occurs in
the relative phase between the $s$- and $d$--waves. As shown in
Eq.~(20) this phase difference is model independent. If we consider
two Breit-Wigner type resonances for both, $S_{11}(1535)$ and
$D_{13}(1520)$ this phase difference would be rather constant as both
resonances are very close in their energy position and, furthermore, have
a similar resonance width. From the fact that the $S_{11}$ is a 
bit higher in energy, the phase difference $\Phi_0 - \Phi_2$
should be negative as is shown in the figure as the dotted line.
\begin{figure}[htbp]
\centerline{\psfig{file=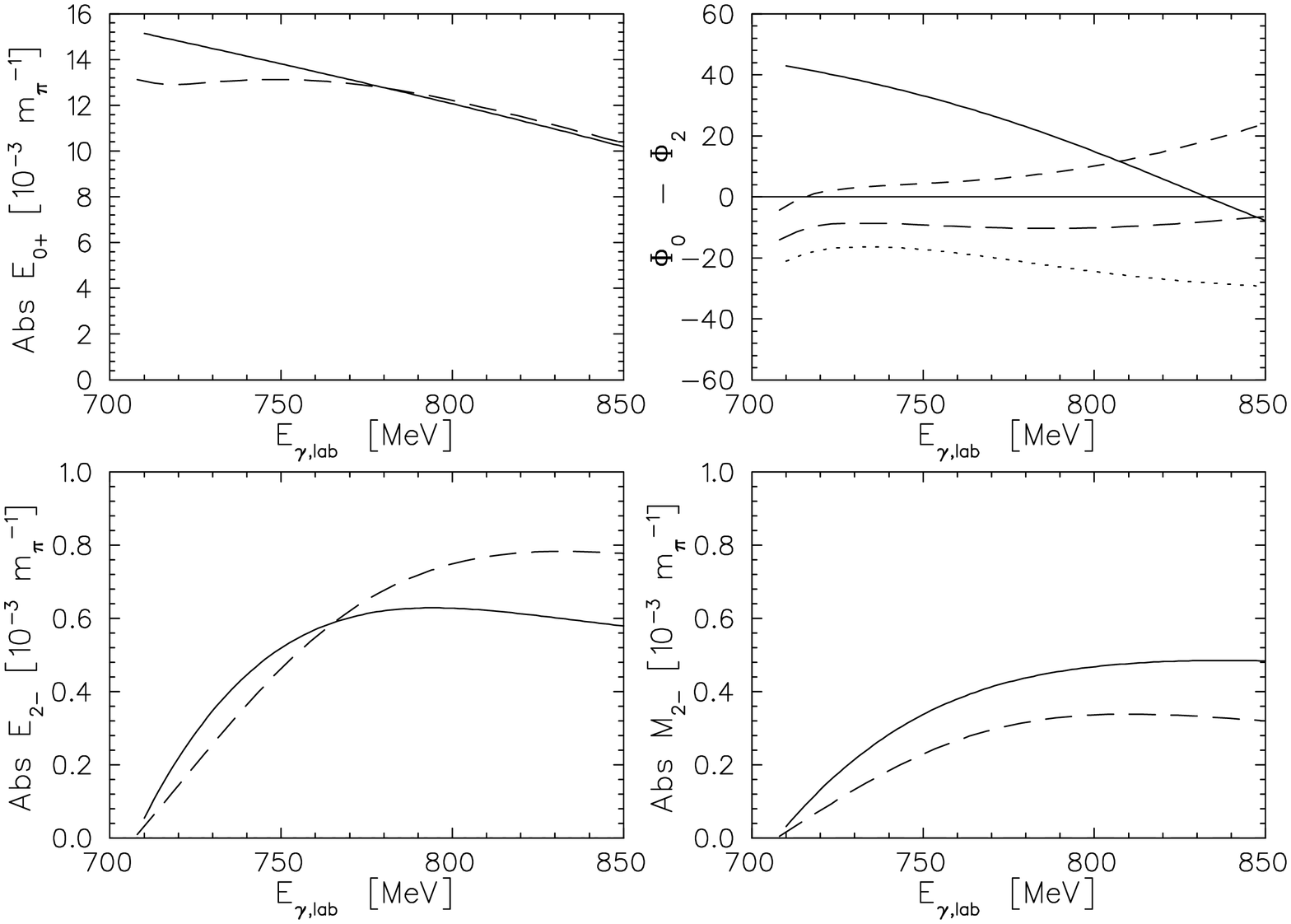,width=12.5cm,angle=90,silent=}}
\caption[thanks]{\label{fig4} Result of the multipole analysis for
  $s$- and $d$- waves. The solid lines show the result of the fit.
  The short and long dashed lines are obtained from the isobar model.
  In the upper right figure we compare the phase difference of our fit
  with the isobar model. The short and long dashed curves show the
  difference obtained with the $E_{2-}$ and $M_{2-}$, respectively.
  The dotted line is the difference of two Breit-Wigner forms.
}
\end{figure}

>From the above analysis we conclude that this completely unexpected 
discrepancy is directly connected to the node structure of
the target asymmetry. Without a node or with a node but an
$e$--coefficient of opposite sign, the phase difference would be much
smaller and closer to our model calculations.

\section*{ETA PHOTOPRODUCTION AT HIGHER ENERGIES}

The most remarkable fact of eta photoproduction in the low energy
region is the strong dominance of the $S_{11}$ channel. Whether it
occurs from a $N^*$ resonance, which is the most likely case, or from
different mechanism is a very interesting question and subject of many
ongoing investigations. In the experiment it shows up as a flat
angular distribution and only very precise data can observe some tiny
angular modulation as found by the Mainz experiment \cite{Krus95}. At
Bonn, angular distributions of the differential cross section have
been measured up to $1.15$~GeV \cite{Bock97} with no evidence for a
break-down of the $s$--wave dominance. Therefore, we can speculate
that this dominance continues up to even higher energies.
Theoretically, this could be understood in terms of very small
branching ratios for nucleon resonances into the $\eta N$ channel. For
all resonances except the $S_{11}(1535)$ the branching ratio is below
1\%, or in most cases even below 0.1\%. In the case of the
$D_{13}(1520)$ resonance
% with a ratio of $(0.1 \pm 0.2)\%)$\cite{PDG96}
this ratio is also assumed around 0.1\%, however, an average number is
no longer quoted in the Particle Data Tables \cite{PDG96}. Only
branching ratios for the two $S_{11}$ resonances remain. As we have
shown in the last Section, the photon asymmetry is a very sensitive
probe for even tiny branching ratios such as  the $D_{13}$ resonance.

In the following, we demonstrate that this is especially the case for
nucleon resonances with strong helicity $3/2$ couplings $A_{3/2}$.  In
Table 1 we list all entries for $N^*$ resonances with isospin $1/2$.
From this table one finds the $D_{13}$ as the strongest candidate
to show up in the
photon asymmetry.  However, other resonances include the
 $F_{15}(1680)$ which plays an important role in pion
photoproduction and, furthermore, the $F_{17}(1990)$ and the
$G_{17}(2190)$ that are less established in photoproduction reactions.
Furthermore, since these numbers are determined from data in the
pion photoproduction channel, surprises in the eta photoproduction
channel are not only possible but indeed very likely.
% =========================== TABLE ==================================== 
\begin{table}[htbp]
    \caption{Photon couplings and multipolarities for $N^*$ Resonances
      with helicity $3/2$ excitation. The numbers are taken from
      PDG96\protect\cite{PDG96}, average numbers above and
      single quoted numbers (less certain) below the horizontal line.}
  \begin{center}
    \leavevmode
    \begin{tabular}{ccc}
    $N^*$ Resonance & $A_{3/2} [10^{-3}GeV^{-1/2}]$ & Multipoles \\
    \hline\hline
    $D_{13}(1520)$ & $+166\pm 5$  & $B_{2-}=E_{2-}+M_{2-}$ \\
    $D_{15}(1675)$ & $+15\pm 9$   & $B_{2+}=E_{2+}-M_{2+}$ \\
    $F_{15}(1680)$ & $+133\pm 12$ & $B_{3-}=E_{3-}+M_{3-}$ \\
    $D_{13}(1700)$ & $-2\pm 24$   & $B_{2-}=E_{2-}+M_{2-}$ \\
    $P_{13}(1720)$ & $-19\pm 20$  & $B_{1+}=E_{1+}-M_{1+}$ \\
    \hline
    $F_{17}(1990)$ & $+86\pm 60$  & $B_{3+}=E_{3+}-M_{3+}$ \\
    $D_{13}(2080)$ & $+17\pm 11$  & $B_{2-}=E_{2-}+M_{2-}$ \\
    $G_{17}(2190)$ & $81 - 180$   & $B_{4-}=E_{4-}+M_{4-}$ \\   
    \end{tabular}
    \label{tab:res}
  \end{center}
\end{table}

Assuming $S$--wave dominance and therefore $F_1$--dominance in the
amplitude we can derive a general expression for the photon asymmetry,
\begin{eqnarray}
\Sigma(\theta) & = & -\sin^2\theta\,\, 
\mbox{Re}\big[F_1^* F_4\big]/R_T^{00}\, ,\\
& = & \sin^2\theta\,\, \mbox{Re}\bigg[E_{0+}^* \sum_{\ell \ge 2}
(B_{\ell -}+B_{\ell +}) P_\ell''(\cos\theta) \bigg] /R_T^{00}\, \\
\end{eqnarray}
with $B_{\ell-}=E_{\ell-}+M_{\ell-}$ and
$B_{\ell+}=E_{\ell+}-M_{\ell+}$.
Both multipole combinations are helicity $3/2$ multipoles and for
resonance excitation they are proportional to the photon couplings 
$A_{3/2}$. The helicity $1/2$ couplings $A_{1/2}$ do not enter here,
they appear in the differential cross section and in the recoil
polarization, e.g. as $A_{2-}=(3M_{2-}-E_{2-})/2$. Explicitly, we
obtain up to $\ell_{max}=4$
\begin{eqnarray}  
\Sigma(\theta) & = & \frac{\sin^2\theta}{|E_{0+}|^2}\, \mbox{Re}
\Big\{E_{0+}^*\Big[3(B_{2-}+B_{2+})-\frac{15}{2}(B_{4-}+B_{4+})
\nonumber \\
& + & 15(B_{3-}+B_{3+})\cos\theta 
+ \frac{105}{2}(B_{4-}+B_{4+})\cos^2\theta \Big] \Big\}\, .
\end{eqnarray}
In Fig. 5 we demonstrate how such interferences of higher resonances
with the $S_{11}$ channel could show up in the photon asymmetry. Even
if two small resonances of different multipolarity are excited in the
same energy region they will produce a clear signal that will
eventually allow determining $\eta$ branching ratios down to values
well below $0.1\%$.
\begin{figure}[tbp]
\centerline{\psfig{file=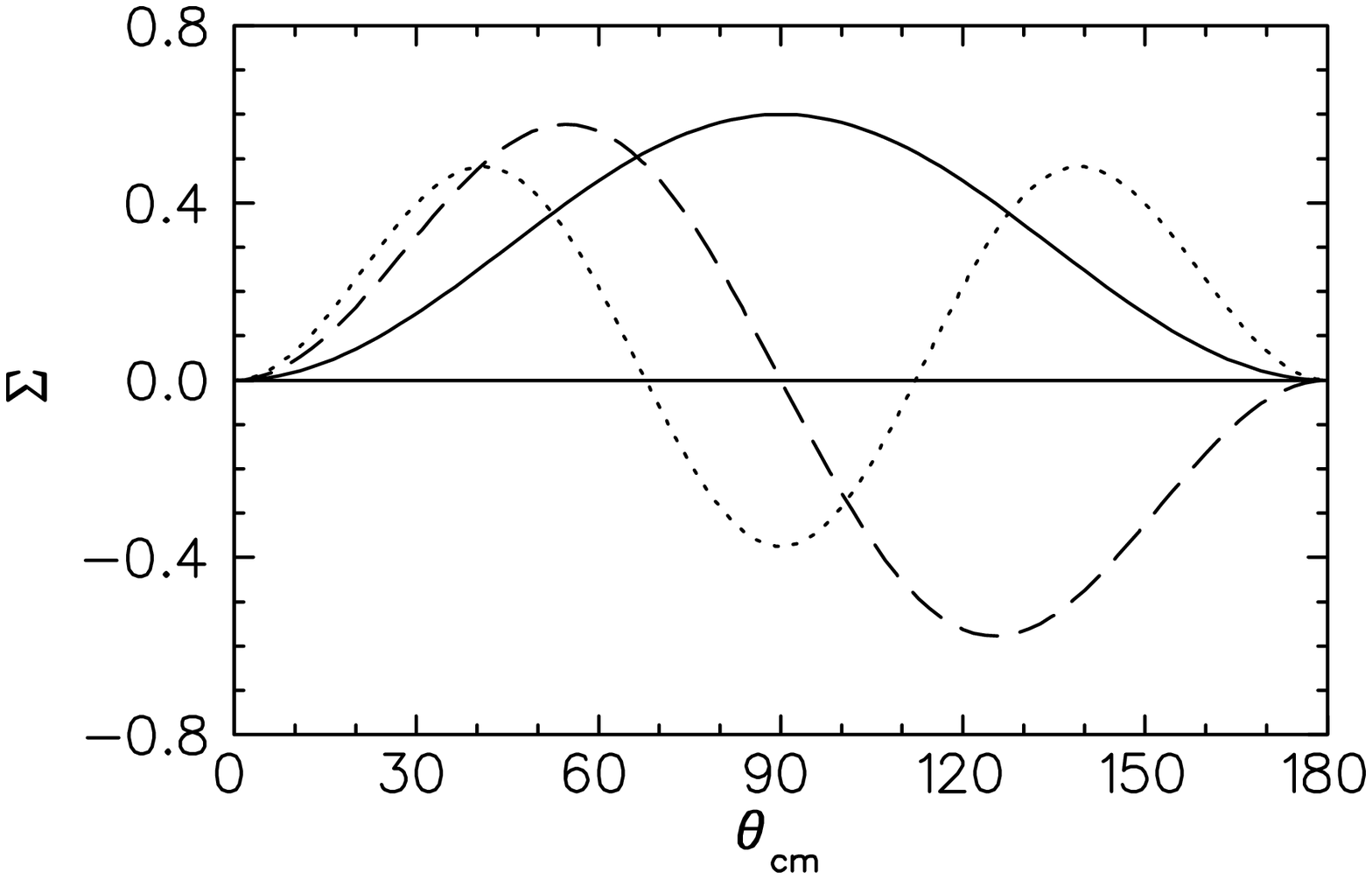,width=7.5cm,silent=}
%\hspace{0.1cm}
\psfig{file=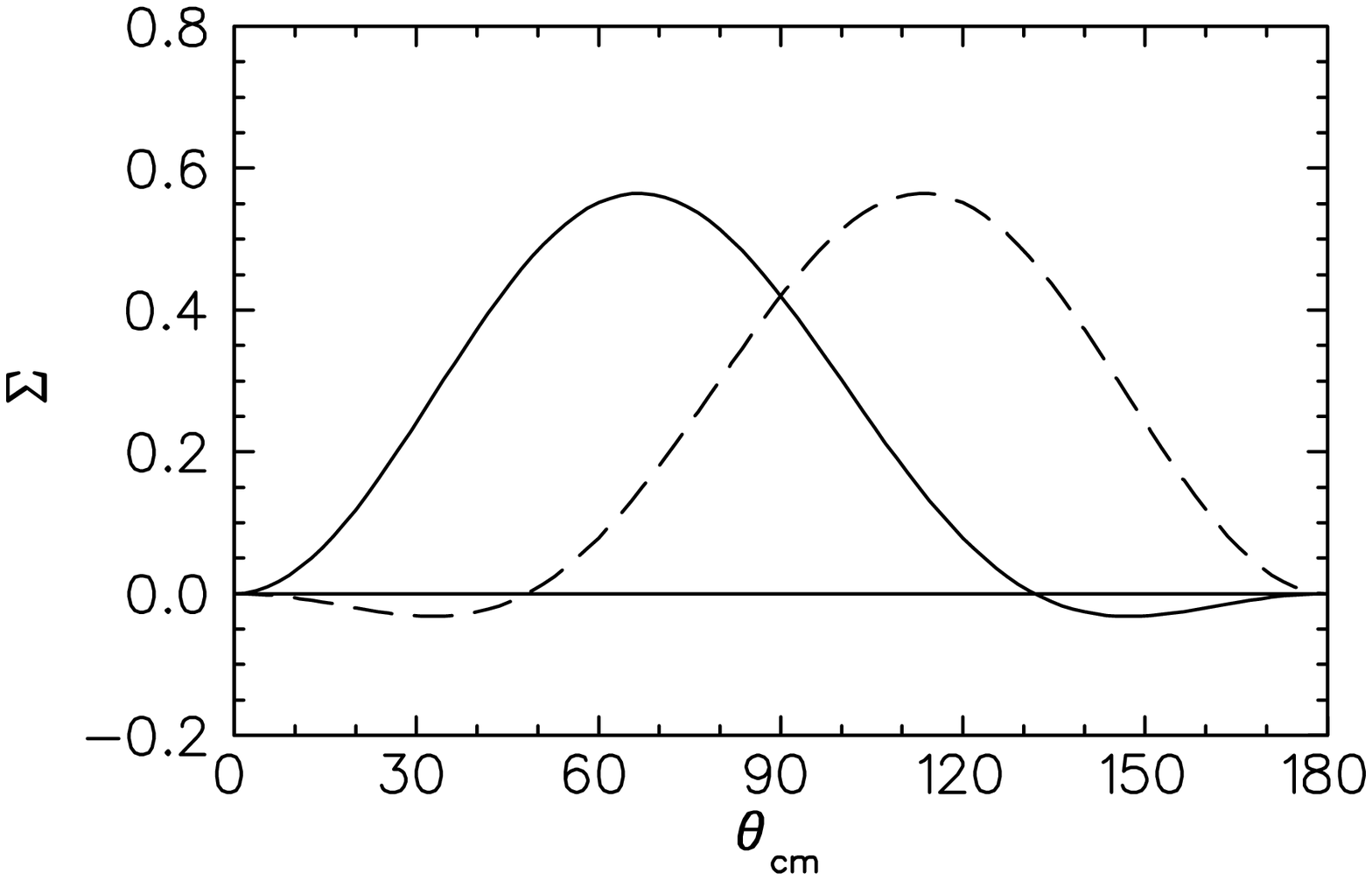,width=7.5cm,silent=}}
\caption[thanks]{\label{fig5} Possible signatures of $N^*$ resonances
  in the photon asymmetry of eta photoproduction. The solid, dashed
  and dotted lines in the left figure show the angular distributions
  for the interference of the dominant $S_{11}$ channel with an
  isolated $D$--, $F$, or $G$--wave, respectively. On the right, the
  situation of two resonances in the same energy region is
  demonstrated for a ($D_{13}$, $F_{15}$) pair (solid curve) and a
  ($D_{13}$, $F_{17}$) pair (dashed curve). Opposite signs are also
  possible if the photon or eta couplings of the resonances obtain a
  negative sign, see Table 1. }
\end{figure}

\section*{SUMMARY}

We have demonstrated that polarization observables are a powerful tool
in analyzing individual resonances in the eta photoproduction channel.
The strong dominance of the $S_{11}$ channel allows a much easier
analysis compared to pion photoproduction. Furthermore, the
nonresonant background in eta physics appears to be small due to a
very weak coupling of the eta to the nucleon.  A combined analysis of
differential cross section, photon asymmetry and target polarization
allows a determination of $s$-- and $d$--wave multipoles. The target
polarization measured at Bonn reveals an unexpected phase shift
between the $S_{11}$ and $D_{13}$ resonances that could lead to the
conclusion that either of these resonances, perhaps the $S_{11}$, is
heavily distorted or is even a completely different phenomenon, as
frequently speculated. The new experiments therefore add another piece
to the eta puzzle that makes the field of eta physics so exciting.

\bibliographystyle{unsrt}

\end{document}